\documentclass{article}\usepackage[]{graphicx}\usepackage[]{color}
\makeatletter
\def\maxwidth{ %
  \ifdim\Gin@nat@width>\linewidth
    \linewidth
  \else
    \Gin@nat@width
  \fi
}
\makeatother

\definecolor{fgcolor}{rgb}{0.345, 0.345, 0.345}

\usepackage{framed}
\makeatletter
 {\par\unskip\endMakeFramed%
 \at@end@of@kframe}
\makeatother

\definecolor{shadecolor}{rgb}{.97, .97, .97}
\definecolor{messagecolor}{rgb}{0, 0, 0}
\definecolor{warningcolor}{rgb}{1, 0, 1}
\definecolor{errorcolor}{rgb}{1, 0, 0}

\usepackage{alltt}
\usepackage[utf8]{inputenc}
\usepackage{amsmath,amssymb,indentfirst,epsfig,url}
\usepackage{makeidx} 
\usepackage{graphicx}
\usepackage{subcaption} 
\usepackage{multicol} 
\usepackage{multirow} 
\usepackage{hyperref} 
\hypersetup{colorlinks=true, linkcolor=red, citecolor=blue, filecolor=black, urlcolor=green} 
\usepackage{verbatim} 
\usepackage[affil-it]{authblk} 
\usepackage{natbib} 
\bibpunct[; ]{(}{)}{,}{a}{,}{;} 

\newtheorem{definition}{Definition}
\providecommand{\keywords}[1]{\textbf{\textit{Keywords: }} #1}

\IfFileExists{upquote.sty}{\usepackage{upquote}}{}
\begin{document}
	\baselineskip = 5.7mm  
	
	\title{Stress Testing Network Reconstruction via Graphical Causal Model}
	
	\author[1]{Helder Rojas\thanks{Electronic address: \texttt{hmolina@santander.com.br}}}
	\affil{\small{ Institute of Mathematics and Statistics, University of S\~ao Paulo -- IME-USP, Brazil}}
	
	\author[2]{David Dias\thanks{Electronic address: \texttt{davdias@santander.com.br}}}
	\affil{\small{Institute of Mathematical and Computer Sciences, University of S\~ao Paulo -- ICMC-USP, Brazil}}
	\affil[1,2]{\small{Santander Bank, Brazil}}
	
	\date{} 

	\maketitle
	
	\begin{abstract}

An resilience optimal evaluation of financial portfolios implies having plausible hypotheses about the multiple interconnections between the macroeconomic variables and the risk parameters. In this paper, we propose a graphical model for the reconstruction of the causal structure that links the multiple macroeconomic variables and the assessed risk parameters, it is this structure that we call Stress Testing Network (STN). In this model, the relationships between the macroeconomic variables and the risk parameter define a "relational graph" among their time-series, where related time-series are connected by an edge. Our proposal is based on the temporal causal models, but unlike, we incorporate specific conditions in the structure which correspond to intrinsic characteristics this type of networks. Using the proposed model and given the high-dimensional nature of the problem, we used regularization methods to efficiently detect causality in the time-series and reconstruct the underlying causal structure. In addition, we illustrate the use of model in credit risk data of a portfolio. Finally, we discuss its uses and practical benefits in stress testing.
\\
\\	
\end{abstract}

\section{Introduction}

	Stress testing resurfaced as a key tool for financial supervision after the 2007-2009 crisis; before the crisis these tests were rarely exhaustive and rigorous and financial institutions often considered them only as a regulatory exercise with no impact at capital. The lack of scope and rigor in these tests played a decisive role in the fact that financial institutions were not prepared for the financial crisis. Currently, the regulatory demands on stress tests are strict and the academic interest for related issues has increased considerably. Stress tests were designed to assess the resilience of financial institutions to possible future risks and, in some cases, to help establish policies to promote resilience. Stress tests generally begin with the specification of stress scenarios in macroeconomic terms, such as severe recessions or financial crisis, which are expected to have an adverse impact on banks. A variety of different models are used to estimate the impact of scenarios on banks’ profits and balance sheets. Such impacts are measured through the risk parameters corresponding to the portfolios evaluated. for a more detailed on the uses and stages of the stress tests, see \cite{dent2016stress} and \cite{henry2013macro}.
	
One of the main objectives in stress testing is to explain the behavior of the financial risk parameters of banking institutions based on macroeconomic variables. In general, this task is not simple, given that the procedures for the selection of macroeconomic variables basically include heuristic processes, based on correlations whose objective is to select a subset of macroeconomic variables and at the same time comply with the assumptions of model used. Something important to be taken into account is that the financial risk parameters are developed within a fully interconnected macroeconomic context, this can be easily corroborated since since the absolute value of the bivariate correlations between the risk parameters and each macroeconomic variable are usually close to one. This reinforces the hypothesis that the temporal evolution of risk parameters is not only related to a subset of macroeconomic variables, but rather is part of a highly interconnected complex system. Therefore, as discussed below, in highly interconnected contexts, conventional correlation-based procedures is not the best way to select candidate variables for econometric models.

\begin{figure}[h!]
		\centering
		\begin{subfigure}[b]{0.35\linewidth}
			\includegraphics[width=\linewidth]{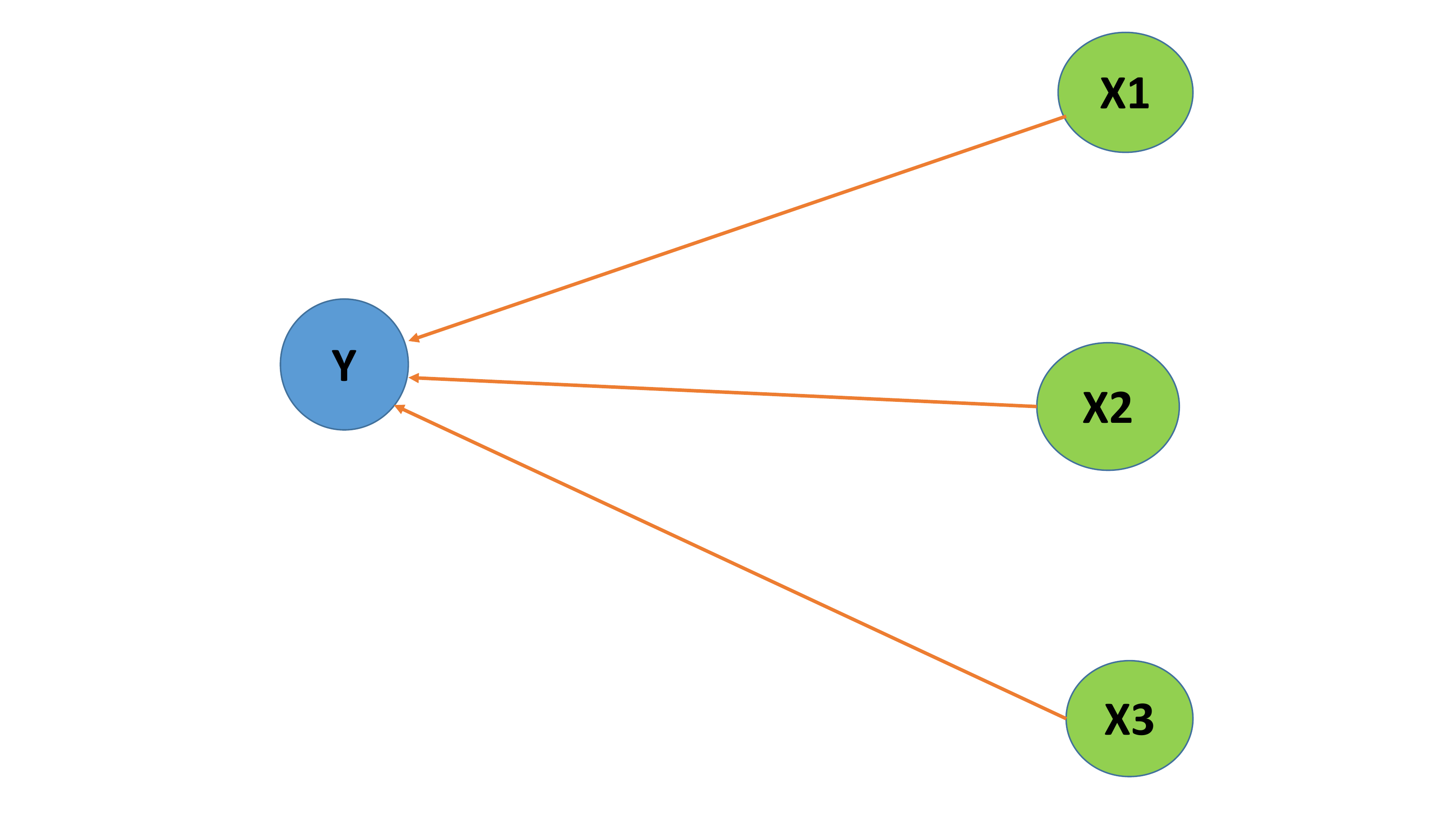} 
		\end{subfigure}
		\begin{subfigure}[b]{0.35\linewidth}
			\includegraphics[width=\linewidth]{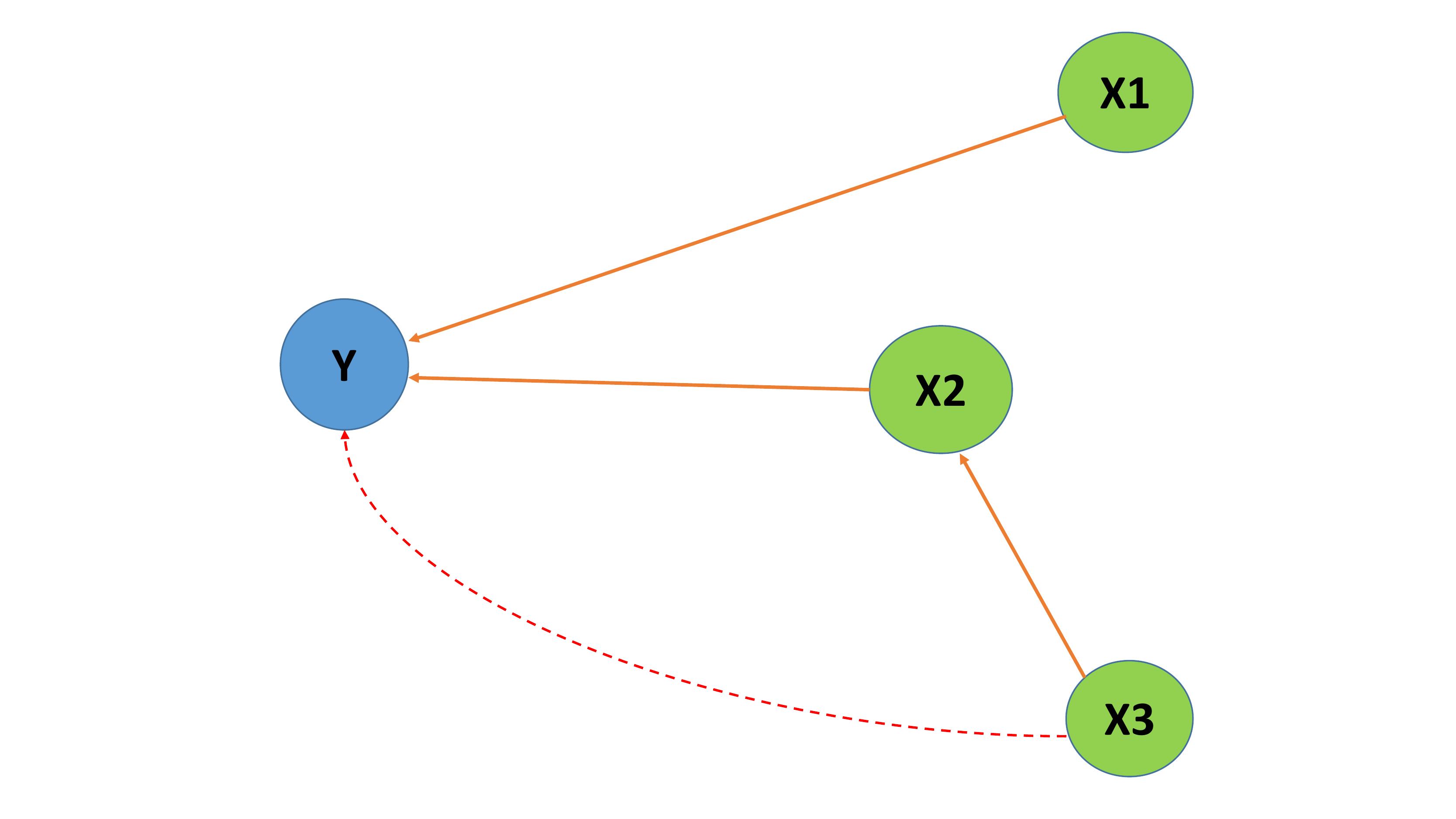}
		\end{subfigure}
		\begin{subfigure}[b]{0.35\linewidth}
			\includegraphics[width=\linewidth]{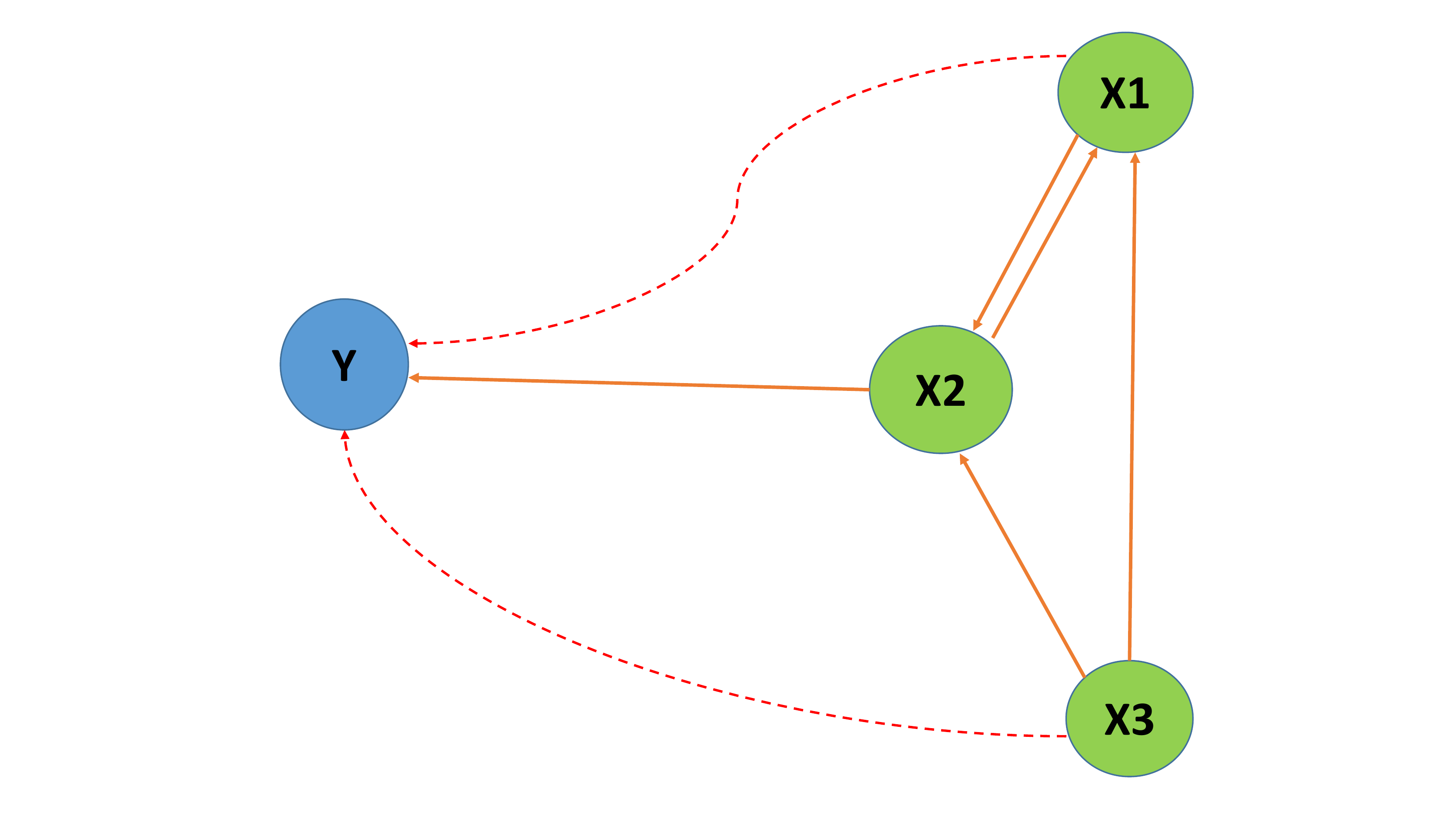}
		\end{subfigure}
    \caption{Configurations contemplated in the hypothetical example.}
     \label{graphs_examples}
	\end{figure}

To better understand the above we will discuss the following hypothetical example. Let $Y$ be a risk parameter that we present to model, in addition, be $X1$, $X2$ and $X3$ macroeconomic variables candidates for the model, selected by conventional procedures. Assuming that we are using a linear model to represent the relationship between the variables, which would imply that a linear combination of $X1$, $X2$ and $X3$ would reasonably explain the behavior of $Y$. The model will determine a type of causal relationship between the variables, in which the macroeconomic variables directly impact the dynamic behavior of the risk parameter, and in turn, the macroeconomic variables do not interact with each other, see upper left subfigure of the Figure \ref{graphs_examples}. Based on our experience in the analysis of many financial portfolios, this type of simplified description of the relationships between the variables, leads to unstable models as a direct effect of not to condense the genuine causal relationships between macroeconomic variables and risk parameters .
\newline

As we stated before, a more realistic scenario is to assume that the variables are part of a highly interconnected system, to assume otherwise would imply a fragmented macroeconomic context which is neither coherent nor economically nor empirically. A more realistic scenario corresponds to that represented by the upper right subfigure of Figure \ref{graphs_examples}, in this configuration $X3$ causes $X2$ and then $X2$ causes $Y$, in addition, $X1$ causes $Y$. In this second configuration, both $X1$ and $X2$ have a direct causal relationship towards $Y$, but in case $X3$, this only influences $Y$ through $X2$, which is configured as an indirect causal relationship that is possibly manifested in a high induced correlation between $X3$ and $Y$ (dotted lines). From the subfigure, we deduce that if we want to build a model for the risk parameter $Y$ that condenses direct causal relationships, we must consider only $X1$ and $X2$, but if we choose a model that contemplates the three variables simultaneously, the interaction between $X2$ and $X3$, product of interdependence, should also be considered in the model. The lower center subfigure of Figure \ref{graphs_examples} shows a scenario of greater interconcection between the variables, where only $X2$ has a direct causal relationship to $Y$, however, the three variables are highly correlated with Y as an indirect result of the interconnections (dotted lines). Therefore, it is clear that building models in stress testing using criteria for selecting variables based on correlations will rarely give good results. Note that this analysis assumes a simple linear model, but the same arguments be valid for models with autoregressive terms, latent hidden states or even more general models. In addition, the problem of identification of causal relationships is more interesting and complex with the inclusion of many more macroeconomic variables, which is known in the literature as high-dimensional causal discovery problems that is the subject of this paper.

We propose to use a graphic description of the causal structure between the macroeconomic variables and the risk parameters as a tool for the construction of models in stress testing and thus guarantee an optimal evaluation of the resilience of the risk parameters. Since this structure is underlying the data and not directly observable, our main goal in this paper is to reliably reconstruct this structure, i.e., reconstruct the causal structure that links a risk paramater and the multiple macroeconomic variables, this structure is what we refer to as Stress Testing Network (STN). The notion of causality that we use in this paper is of temporal sense, i.e., the cause must precede the effect, therefore,  it is consistent with the "Granger Causality" \cite{granger1980testing}. In this paper we propose a graphical model for the STNs in which the relationships between the macroeconomic variables and the risk parameter define a "relational graph" among their time-series, where related time-series are connected by an edge. Our proposal is based on the temporal causal models, see \cite{eichler2012graphical}, \cite{lozano2009grouped} and \cite{mukhopadhyay2006causality}, but unlike, we incorporate specific conditions in the structure which correspond to intrinsic characteristics of the STNs. Following the proposed model and given the high-dimensional nature of the problem, we used regularization methods to efficiently detect Granger's causality in the time series and reconstruct the underlying causal structure. We also illustrate the use of model in credit risk data of a portfolio.

\subsection{Motivation} 

Our original interest arises from our study of transmission of macroeconomic shocks to risk parameters, see \cite{rojas2018transmission}. In that work we describe the main characteristics of the transmission processes to incorporate them into an extended family of models called general transfer function models. One of the most complete representations of the model assumes the existence of a dynamic latent state $E_{t}$ to which the macroeconomic shocks are transferred, the joint transference of the shocks to the latent state occurs in an additive and linear way, i.e., a linear combination of the macroeconomic variables which implies that these should be conditionally independent. The transfer of macroeconomic shocks to the latent state is gradual and persistent, thus generating endogenous behavior to the latent state that is incorporated into the model through the autoregressive term $\phi_{t} E_{t-1}$. The dynamic character of $\phi_ {t}$ is due to the metastable nature of the transmission. Finally, macroeconomic shocks are transferred from the latent state to the risk parameter with the incorporating an ambient noise that is assumed Gaussian. An implication this model that is interesting and consistent with the data, is the fact the risk parameter inherits the endogenous behavior of the latent state. 
\newpage
This model successfully condenses many of the properties observed in the data, but for simplicity the absence of interdependencies between macroeconomic variables is assumed since the latent state transfers a linear combination of shocks without taking into account any possible interaction between macroeconomic variables. Therefore, we have two options: using macroeconomic variables that do not present direct causality or including an interaction structure between the variables in the dynamics of the latent state. For either case, it is necessary to know the causal structure that links all the variables, as this structure is not observable, but rather underlying the data, it was the motivation to consider the reconstruction problem of causal network. It is important to mention that knowing the causality relationships between the variables allows us to have an overview of the interconnections, which is useful when building models, therefore, although the causal network reconstruction was motivated by transfer function models, its utility is independent of the model evaluated.
	\subsection{Outline} 
	
    The paper is organized as follows. In Section \ref{s:ttn} we describe the characteristics of the STNs. In section \ref{s:case} we present a case study. Section \ref{s:m} we present the proposed graphical model for network reconstruction and the procedure for estimating graphical model. In section \ref{s:ac} we present results of the model applied to study case. Finally, in section \ref{s:stn_in_stm} we discuss how STN can be used in stress testing modeling.

\section{Stress Testing Networks}\label{s:ttn}
	
The risk of possible losses in the portfolios of financial institutions is monitored through what are known as risk parameters. Each risk dimension (for example credit risk, interest rate risk, market risk and others) is monitored by different risk parameters. The main objective in stress testing is to infer the behavior of the risk parameters in hypothetical downturn macroeconomic scenarios, based on historical series of the risk parameter and macroeconomic variables, therefore, as discussed above it is important to understand the causal structure that links them. As mentioned above, this causal structure between a risk parameter and the macroeconomic variables is what we call STN.

The risk parameters are calculated individually for each financial institution, in addition, are calculated for each predefined portfolio within the financial institution, therefore, the aggregation level of this variable is much lower compared to the macroeconomic variables. This fact leads to the absence of a causality relationship between risk parameters to macroeconomic variables, i.e., the risk parameters individually do not significantly impact the macroeconomic context. This absence of causality is included as conditions on graphic structure which we propose for reconstruction of the STNs,  see section \ref{Model}. Another common feature in this networks is their sparse nature, i.e.,  not all variables are connected by causality links. This fact is due to that, although the variables have a high correlation, induced for indirect causal relationships, the causality relations underlying these variables represent a system of less complexity than that observed in the correlations. Therefore, sparsity is assumed in the reconstruction of the STNs, see section  \ref{Estimation}.

In general, the macroeconomic variables considered for the reconstruction of STNs, both paramteros of risk as the multiple macroeconomic variables, follow random walks within the period in which they are observed simultaneously, which are usually only from 20 to 30 trimonthly observations. This stylized fact in the reconstruction of STNs emphasizes the high-dimensional context in which our problem is configured. It should be noted that all the variables considered in the reconstruction are oberved in level, avoiding transformations that can exclude relevant information about their dynamic behaviors. Only in the case of risk parameters that are considered monotonic transformations, which represent an isomorphism between the domain of the parameter and the real line, so that they can be analyzed using linear models.

As will be seen in more detail in section \ref{s:m}, as the problem of reconstruction of STNs is configured in high-dimension contexts, regularization methods, Lasso and Elastic-Net, will be used to infer the proposed graphical model which will provide us with a approximation of the underlying network to the data. Network reconstruction techniques based on regularization methods have been used successfully in various areas, to see examples in fields such as Biology, Genetics and Climatology, see \cite{liu2010learning}, \cite{chiquet2015contributions}. Although, regularization techniques have been used to select variables in forecasting models in stress testing, see for example \cite{chan2017lasso}, there are no studies available in the literature on network reconstruction in stress testing contexts. However, as already discussed, the practical benefits of using these stress testing techniques are relevant.
	
\section{Case Study: Credit Risk Data}\label{s:case}
	
	Credit risk is a component with great potential to generate losses on its assets and, therefore, has significant effects on capital adequacy. In addition, the credit risk is possibly the dimension of risk with the bigger bank regulation regarding stress tests. The most relevant credit risk parameters to assess resilience are: Probability of Default (PD) and Loss Given Default (LGD). Other risk parameters can be considered, but in general these other parameters have a definition superimposed with the parameters already mentioned. Furthermore, PD and LGD are used explicitly in the calculation of regulatory capital required from banking institutions.

The definition of default and of the parameters mentioned above is given below:
\begin{definition}
	In Stress Testing, it is considered to be the default the borrower who does not fulfill the obligations for determined period of time.
\end{definition}

\begin{definition}
Probability of default (PD) is a financial term describing the likelihood of a default over a particular time horizon (this time horizon depends of the financial institution). It provides an estimate of the likelihood that a borrower will be unable to meet its debt obligations. The most intuitive way to estimate PD is through the Frequency of Observed Default (ODF), which is given by number of bad borrower (or borrower in default) in time t and under total number borrower (number of bad and good borrower) in time t.
\end{definition}

\begin{definition}
Loss Given Default or LGD is the share of an asset that is lost if a borrower defaults. Theoretically, LGD is calculated in different ways, but the most popular is 'Gross' LGD, where total losses are divided by Exposure at Default (EAD). Thus, the LGD is the total debt\% minus recuperate of the debt\%,  i.e., the percentage of debt that was recovered by the financial institution with the borrower's payments.
\end{definition}
For a better understanding of the different parameters of credit risk and their impacts in the calculation of regulatory capital, see \cite{henry2013macro}. Without loss of generality, since they can be applied to other risk dimensions, we focus on the credit risk data, in particular, in the PD of a portfolio monitored from the second quarter of 2009 to the first quarter of 2015, observed quarterly. Macroeconomic variables that are considered for reconstruction, see Appendix \ref{macros},  are also observed in the same period and with the same frequency. The transformation considered for the PD is the logistic.

	\section{Graphical Causal Model}\label{s:m}

           \subsection{Definition Model}\label{Model}

Let $V=\{ X_{1}, X_{2}, \ldots, X_{p}\}$ be the set of variables of interest, i.e., the risk parameter in the first coordinate and  the $p-1$ macroeconomic variables located in the remaining coordinates. The causal structure  can be conveniently represented by a unweighted directed graph $G=(V,E)$, where $V$ is the set of nodes and $E$ is the set of edges in $V \times V$. A pair $(i,j)$, denoted by $i \to j$, is contained in the edge set $E$, if and only if, $X_{i}$ has a causal effect on $X_{j}$  given all the causal effects of remaining variables $X_{V\backslash \{i,j\}}=\{X_{k}; k\in V\backslash \{i,j\}\}$. Let $\mathbf{A}$ denote the adjacency matrix of $ G$, with $A_{ij}=1$ if $i \to j$ and $A_{ij}=0$ otherwise. Note that every configuration not contained in the edge set does not represent a causality relationship. 
Let us denote by $(X_{t})_{t\in \mathbb{N}}$ the $\mathbb{R}^{p}$-valued stochastic process that represents the discrete-time evolution of the variables contained in $V$. Based on empirical facts regarding the response of risk parameters to macroeconomic shocks that show a Markovian dynamic, see \cite{rojas2018transmission}, we assume that the underlying process that generates $ X_ {t} $ is the following way

\begin{equation}\label{gcm}
	(\mathbb{I} -  \Psi)  X_{t}=\Phi X_{t-1}+\textbf{b}+\omega_{t}, \quad \quad t \in \mathbb{N}^{*}
	\end{equation}

where $\Psi=(\psi_{ij})_{i,j\in V}$ and $\Phi=(\phi_{ij})_{i,j \in V}$  are $p\times p$ matrix that incorporate, in the temporal evolution of the process, the contemporaneous effects and the lag effects respectively, the intercept $\textbf{b}$ is a size-$p$ vector and $\omega_{t}$ is a white Gaussian process. Namely, $\omega_{t} \sim GP(0,\textbf{D})$ where is a diagonal matrix such as $\textbf{D}_{ii}=\sigma_{i}^{2}$ and cov$(\omega_{t},\omega_{s})=\textbf{1}_{\{s=t\}}\textbf{D}$ for all $s, t >0$. Moreover, $X_{0} \sim GP(\mu, \Sigma)$, with $\mu$ a size-$p$ vector of means and $\Sigma$ a covariance matrix. Also assume that cov$(X_{t}, \omega_{s})=0$ for all $s>t$: hence, $X_{t}$ is obviously a first-order Markov process.

 Due to the acyclic nature of the causal structure and the inexistence of causal relationships of the risk parameters to macroeconomic variables, it is necessary to establish conditions in (\ref{gcm}) as follows:

	\begin{subequations}\label{cond}
		\begin{align}
		\label{cond1}
		\psi_{ii}&=0, \quad \quad \textrm{for}\quad i=1,\ldots,p\\ 
		\label{cond2}
		\psi_{1j}&=0, \quad \quad \textrm{for}\quad j=2, \ldots, p\\
		\label{cond3}
		\phi_{i1}&=0, \quad \quad \textrm{for} \quad i=1, \ldots,p 
		\end{align}
         \end{subequations}
where (\ref{cond1}) and (\ref{cond2}) correspond to contemporary effects conditions, on the other hand, (\ref{cond3}) corresponds to the lag effects conditions. The nonzero entries in $\Psi$ or $\Phi$ correspond to nonzero entries in $\mathbf{A}$, likewise, the null entries in $\Psi$ and $\Phi$ correspond to null entries in $\mathbf{A}$ . Therefore, inferring $\Psi$ and $\Phi$ is equivalent to reconstructing this graph which is the main issue of this paper.

In this graph, the in-degree of a node $i \in V$ is given by $\mathcal{N}^{in}(i):=\{k\in V\backslash\{i\}: k \to i \}$ which represent the nodes set that have a causal relationship to $i$, while its out-degree $\mathcal{N}^{out}(i):=\{k\in V\backslash\{i\}: i \to k \}$ represents the nodes for which $ i$ has a causal effect. we define also $\mathcal{N}e(i):=\mathcal{N}^{in}(i) \cup \mathcal{N}^{out}(i)$ which represent of nodes set that has a direct causality relationship with $i$. Denote the clouse of $i$ by $\mathcal{C}l(i):=\mathcal{N}e(i) \cup \{i\}$, then $V\backslash \mathcal{C}l(i)$ represent the nodes set that do not have a direct causality relationship with $i$.

         \subsection{Graph Sparse Estimation}\label{Estimation}

 Obtain a estimation of graph, denoted by $\hat{G}=(V,\hat{E})$, is equivalent to estimating the parameters of the equation ($\ref{gcm}$). Remember that, nonzero entries in $\Psi$ or $\Phi$ correspond to nonzero entries in $\mathbf{A}$, likewise, the null entries in $\Psi$ and $\Phi$ correspond to null entries in $\mathbf{A}$. To this end, assume that $X_{t}$ is observed on the time space $t=1, 2 \ldots, T$. Define $p\times T$ matrix $\mathbf{X}=(X_{1}, X_{2}, \ldots, X_{T})$, whose $t^{th}$ column contains the information relative to the $p$ variables at time $t$. Let $\Theta=(\Psi, \Phi)$, which is of dimension $p\times 2p$. Define a $2p\times 1$ vector  $Z_{t}$ as $Z_{t}=(X_{t}^{'}, X_{t-1}^{'})$, and let $\mathbf{Z}=(Z_{1}, \ldots, Z_{T})$, which has dimension $2p\times T$. Finally, define a $p\times T$ matrix $\Omega=(\omega_{1}, \ldots, \omega_{T})$, then Equation ($\ref{gcm}$) may also be expressed as

\begin{equation}\label{objetive_function}
	\mathbf{X}=\mathbf{b}\mathbf{1}^{'}+\Theta \mathbf{Z}+\Omega
	\end{equation}

with $\mathbf{1}$ denoting a $T \times 1$ vector of ones.
\newline
Under a sparsity assumption on $\Theta$, an initial approach to reduce the dimensionality of the parameter space, introduced by \cite{tibshirani1996regression}, is to apply an $L_{1}$-penality (Lasso penalty) to the convex least squares objetive function. Therefore,  the sparse approximation to a Equation ($\ref{objetive_function}$) is given by

\begin{equation}\label{lasso}
	\min_{\Theta: \hspace{0.3em} \psi_{ii}=0, \psi_{1j}=0, \phi_{i1}=0} \Bigg\{ \frac{1}{2}||\mathbf{X}-\mathbf{b}\mathbf{1}^{'}-\Theta \mathbf{Z}||_{F}^{2}+\lambda||\Theta||_{1} \Bigg\}
	\end{equation}

in which $||\ .\ ||_{F}$  is the Frobenius norm, $||\ .\ ||_{1}$ is the $L_{1}$ norm, and $\lambda \geq 0$ is a penality parameter. An $L_{1}$-penalty will induce sparsity in the coefficient matrix $\Theta$, consequently also in the coefficients matrix $\Psi$ and $\Phi$, by zeroing individual entries. Said sparcity in $\Theta$ is controlled with $\lambda$. In this case, an estimate of $\lambda$ is obtained by cross-validation procedure.

The ($\ref{lasso}$) solution can be obtained by convex optimization procedures that are computationally efficient and at the same time work in high-dimensional contexts, for example, see \cite{davis2016sparse} and \cite{tibshirani2015statistical}. Theoretical properties and performance of $L_{1}$-penalty estimates, in similar contexts described here,  are presented in \cite{basu2015regularized}. It is known that solution to penalized formulation is equivalent to solving $p$ independent LASSO problems on each row of $\Theta$, which is an efficient and fairly simple solution to the problem formulated in $(\ref{lasso}$), for more detail see \cite{meinshausen2006high}.

Given that the variables of interest summarize a highly correlated system, in this paper we will consider the elasticity net penalty as a complementary analysis. The Elastic-Net makes a compromise between the Ridge and the Lasso penalty, see \cite{zou2005regularization}; it solves the convex program

\begin{equation}\label{elasticnet}
	\min_{\Theta: \hspace{0.3em} \psi_{ii}=0, \psi_{1j}=0, \phi_{i1}=0} \Bigg\{ \frac{1}{2}||\mathbf{X}-\mathbf{b}\mathbf{1}^{'}-\Theta \mathbf{Z}||_{F}^{2}+\lambda\Big [\frac{1}{2}(1-\alpha)||\Theta||_{2}^{2} + \alpha||\Theta||_{1} \Big ] \Bigg\}
	\end{equation}

in which $||\ .\ ||_{2}$  is the Euclidean norm and $\alpha \in [0,1]$ is a parameter that can be varied, in this case $\alpha=0.5$. The solution ($\ref{elasticnet}$) is obtained with the same procedure described for the problem ($\ref{lasso}$). Note that the restrictions included both Equations, (\ref{lasso}) and (\ref{elasticnet}), correspond to the conditions in (\ref{cond}).

We could use other functions of penalization in the graph reconstruction, see for example \cite{charbonnier2010weighted} and \cite{nicholson2014structured}, but that would imply having an a priori idea of the graph structure which deserves further investigation.

\section{Application: Case study}\label{s:ac}
	
In this section, we illustrate the use of the proposed model in credit risk data that was presented in section \ref{s:case}. We focus  in the PD which was monitored from the second quarter of 2009 to the first quarter of 2015, observed quarterly. Macroeconomic variables that are considered for reconstruction, see Appendix \ref{macros},  are also observed in the same period and with the same frequency.

	\begin{figure}[h!]
		\centering
		\begin{subfigure}[b]{0.45\linewidth}
			\includegraphics[width=\linewidth]{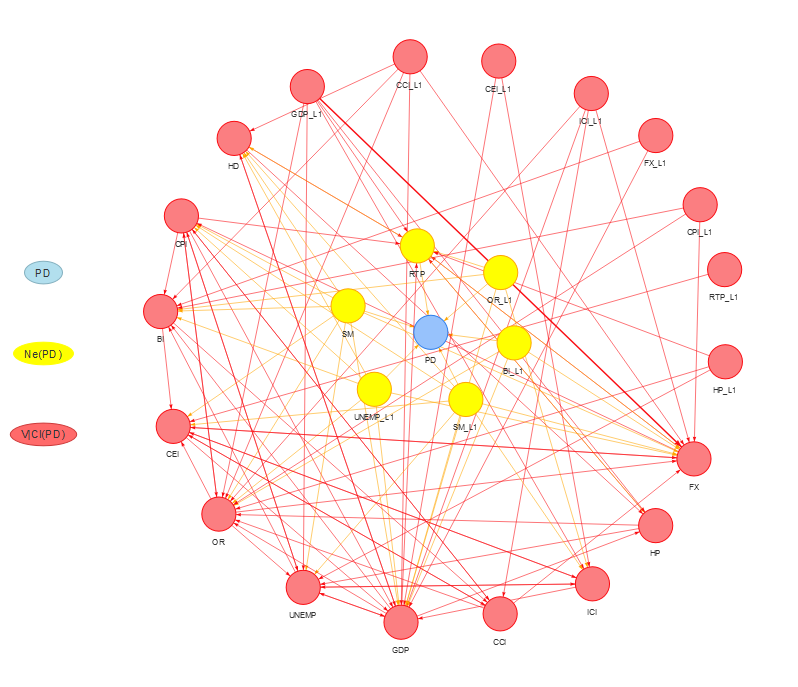} 
		\end{subfigure}
		\begin{subfigure}[b]{0.45\linewidth}
			\includegraphics[width=\linewidth]{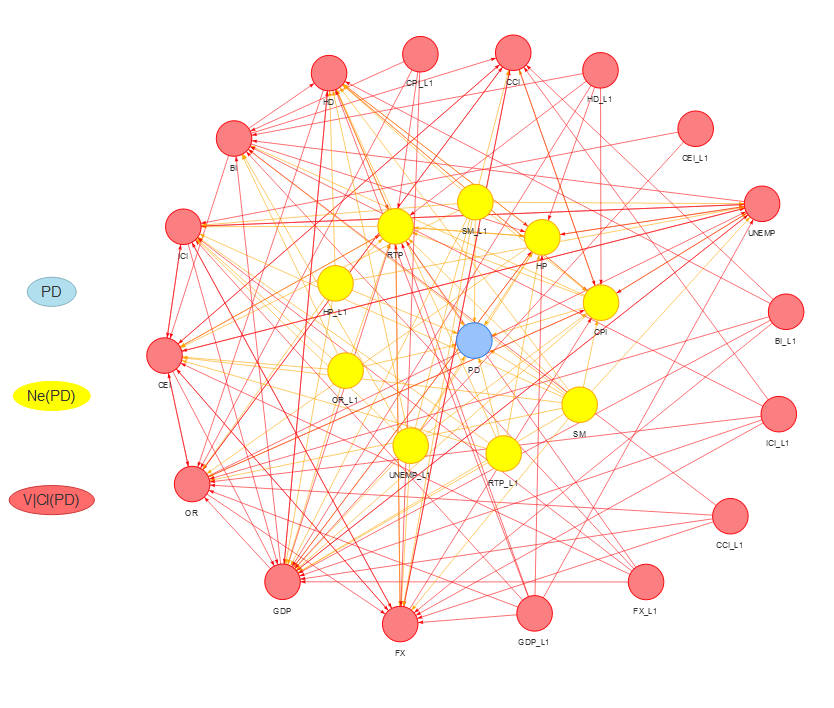}
		\end{subfigure}
    \caption{Directed graph considering the lags as nodes. Left, reconstruction with Lasso penalty. Right, reconstruction with Elastic-Net penalty. }
  \label{graph_full}
	\end{figure}

	\begin{figure}[h!]
		\centering
		\begin{subfigure}[b]{0.32\linewidth}
			\includegraphics[width=\linewidth]{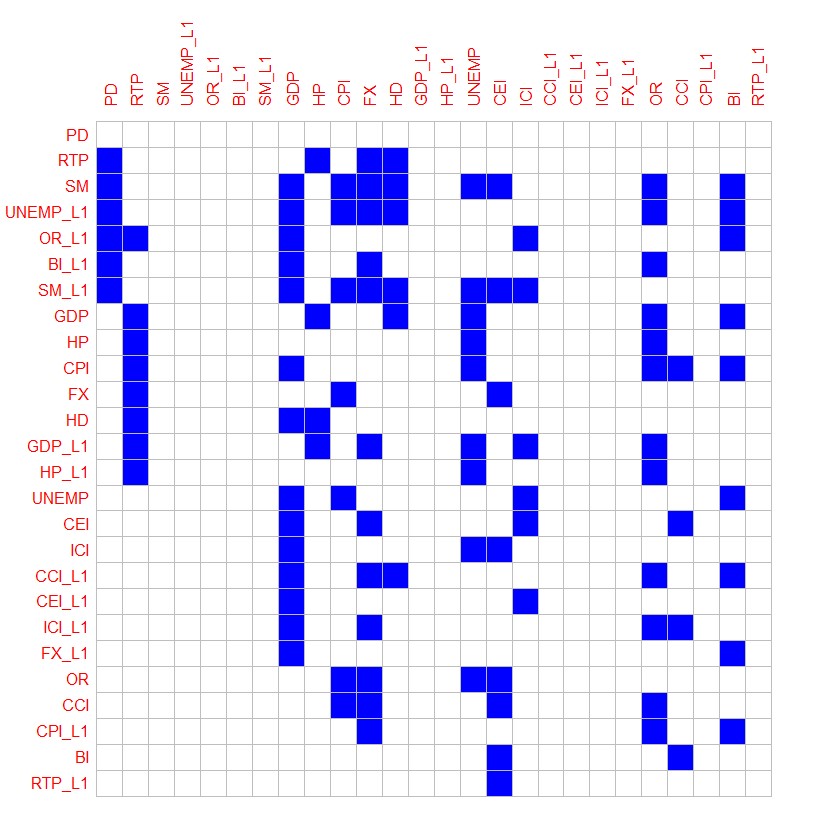} 
		\end{subfigure}
		\begin{subfigure}[b]{0.32\linewidth}
			\includegraphics[width=\linewidth]{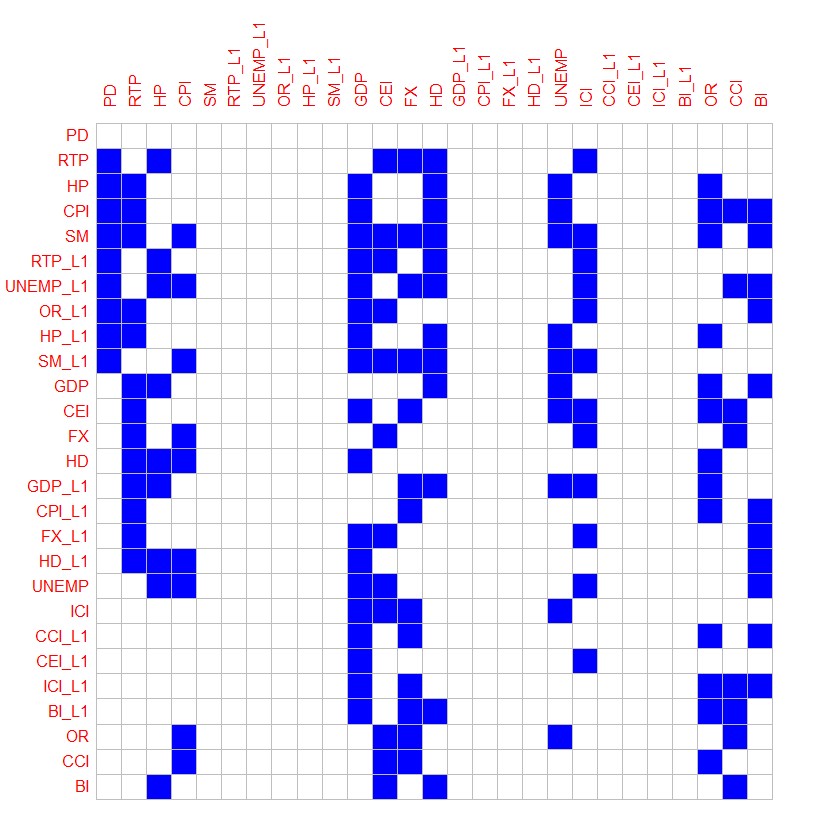}
		\end{subfigure}
    \caption{Adjacent matrix of graph shown in Fig. \ref{graph_full}. Left, reconstruction with Lasso penalty. Right, reconstruction with Elastic-Net penalty.The color blue represent 1 and white 0.}
    \label{matrix_graph_full}
	\end{figure}

\begin{figure}[h!]
		\centering
		\begin{subfigure}[b]{0.49\linewidth}
			\includegraphics[width=\linewidth]{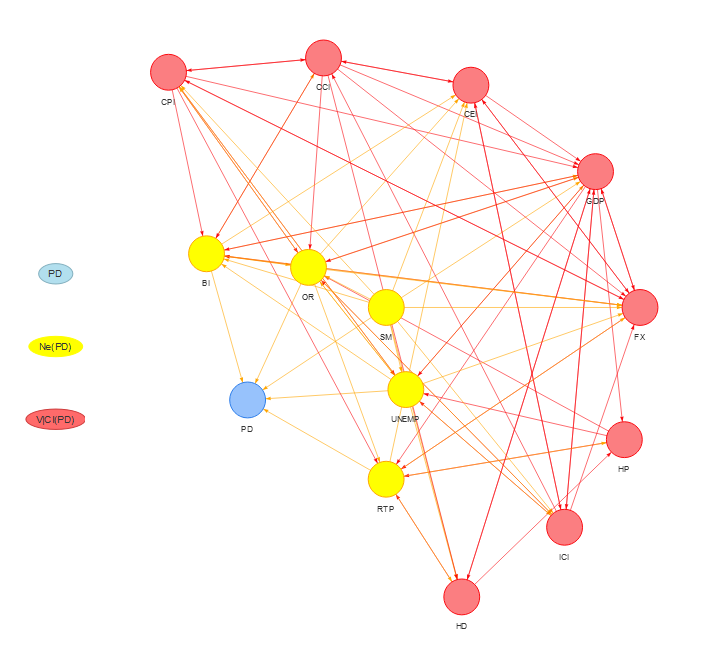} 
		\end{subfigure}
		\begin{subfigure}[b]{0.49\linewidth}
			\includegraphics[width=\linewidth]{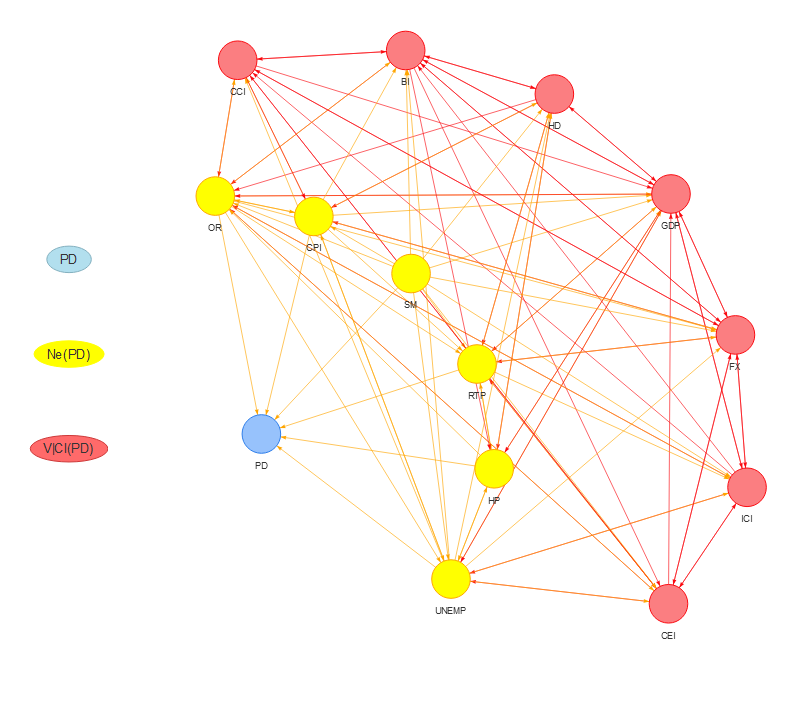}
		\end{subfigure}
    \caption{Estimation of graph, i.e., $\hat{G}=(V,\hat{E})$.  Left, estimation via Lasso penalty. Right, estimation via Elastic-Net penalty.}
  \label{graph_compact}
	\end{figure}

\begin{figure}[h!]
		\centering
		\begin{subfigure}[b]{0.35\linewidth}
			\includegraphics[width=\linewidth]{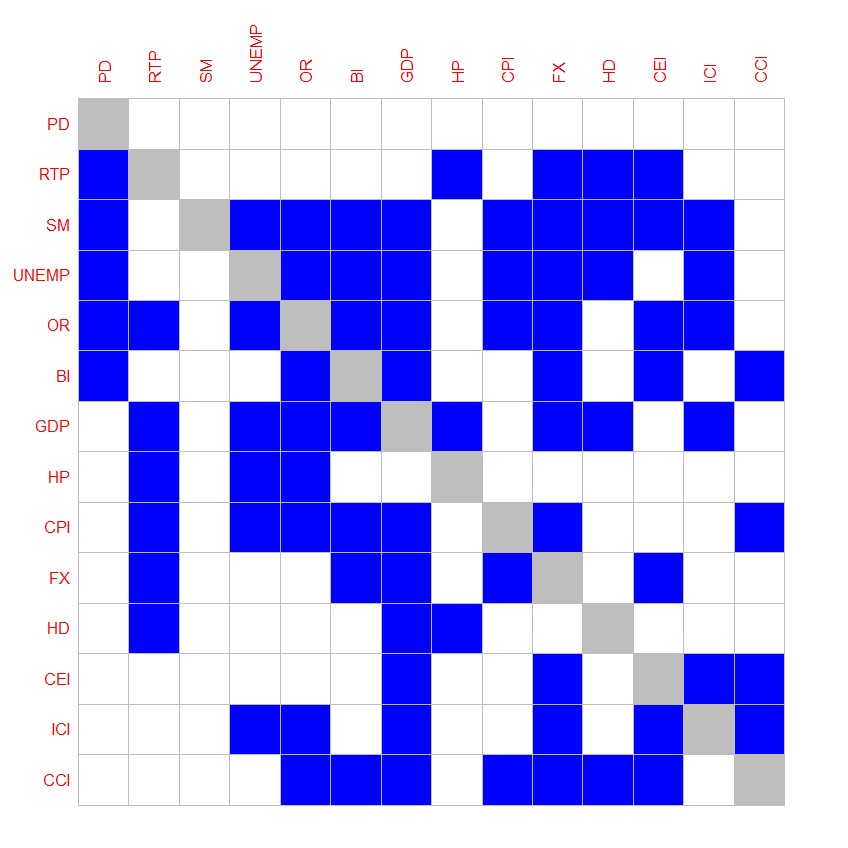} 
		\end{subfigure}
		\begin{subfigure}[b]{0.35\linewidth}
			\includegraphics[width=\linewidth]{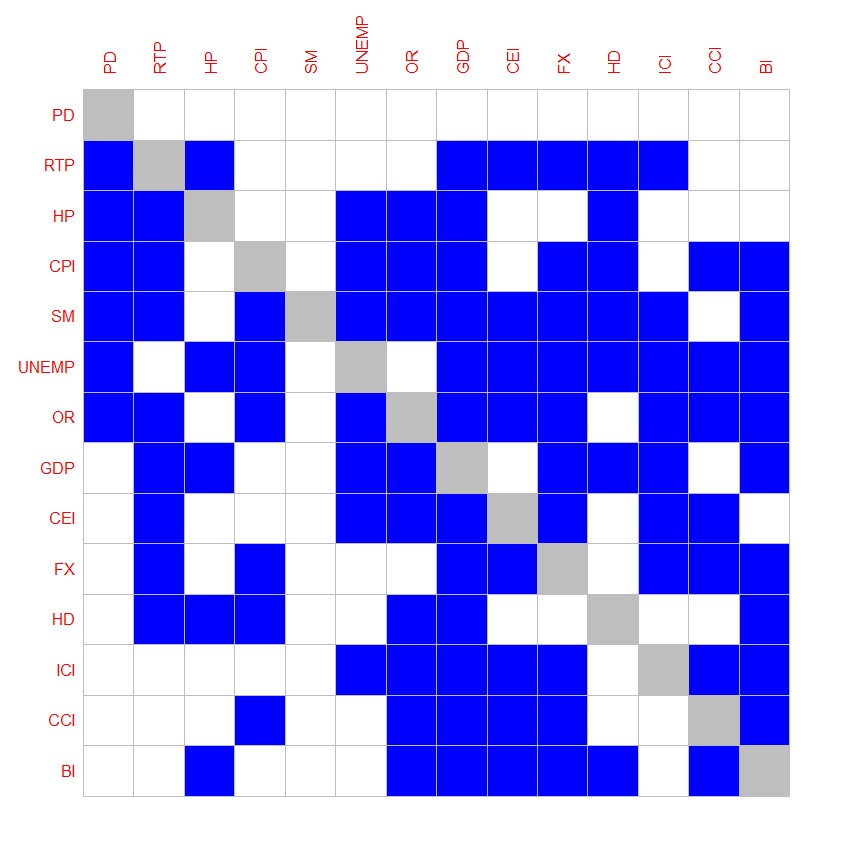}
		\end{subfigure}
    \caption{Adjacent matrix $\mathbf{A}$ of graph shown in Fig. \ref{graph_compact}. Left, reconstruction with Lasso penalty. Right, reconstruction with Elastic-Net penalty. The color blue represent 1 and white 0. Lead represents loops, generated by the variables and their corresponding lags, which represent persistence effects within each variable. }
     \label{Fig:superpositionlags}
	\end{figure}

Figure \ref{graph_full} shows a graph considering the lags of variables as nodes, this graph will be called extended graph. Figure \ref{matrix_graph_full} shows the adiacent matrix of the extended graph. Figure \ref{graph_compact} shows the compact graph, this graph is constructed with the information contained in the extended graph such that if two nodes are linked by at least one axis in the extended graph, regardless of  lag time but maintaining the direction of the relationship, then both are linked in the compact graph. Figure \ref{Fig:superpositionlags} shows the adiacent matrix of the compact graph. Note that the graphs are acyclics, and also, there is no output edge in the PD, this corresponds to the conditions established in Equation \ref{cond}.

The two penalties used in the reconstruction provide us with two relatively different approaches to the interconnections between the variables, the Lasso penalty provides us with a more sparse version than that provided by the Elastic-Net penalty. Something similar happens with the extended and compact graph, the extended graph gives us an approximation of greater temporal detail than the compact graph, but also, the summary and simplification provided by the compact graph is quite useful and parcimonious. The truth is that all these perspectives are very useful when building stress testing models.


\section{STNs in Stress Testing Modeling}\label{s:stn_in_stm}

As we mentioned before, the objective in stress testing is to assess the resilience of the portfolios, consequently a specific objective in stress testing modeling is to model the behavior of risk parameters in terms of macroeconomic variables and to use these dependency relationships to extrapolate the behavior to the risk parameters in hypothetical downturn scenarios. To this end, it is possible to use the characteristics of the transmission process of macroeconomic shocks to the risk parameters to design models that allow us to carry out a better risk assessment, for more details, see \cite{rojas2018transmission}. Since this modeling proposal is based on the information contained in the decay of the shocks of each macroeconomic variable in isolation, multiple effects of shocks can interfere and distort the information contained in the transmission. In addition, we believe that the information contained in the transmission is more clear between variables that maintain a direct causality relationship, this is because, for example, the transimision between variables linked by a third variable suffers more interference.

	\begin{figure}[h!]
		\centering
		\begin{subfigure}[b]{0.4\linewidth}
			\includegraphics[width=\linewidth]{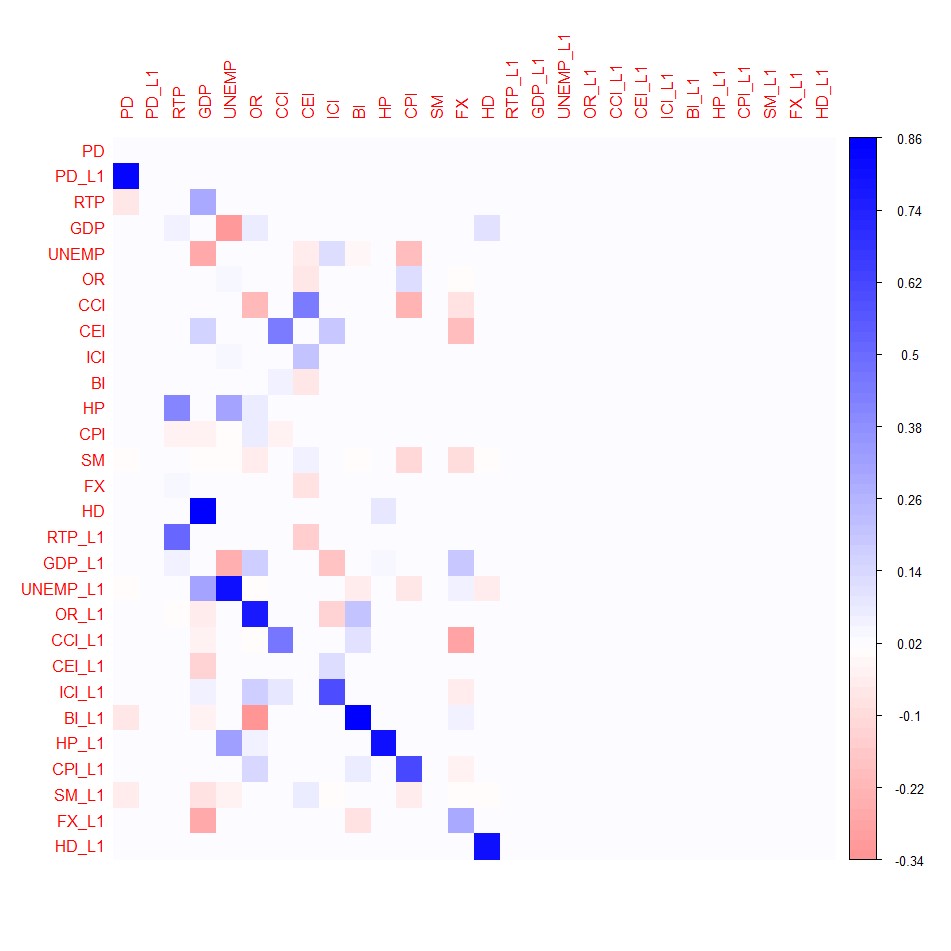} 
		\end{subfigure}
		\begin{subfigure}[b]{0.4\linewidth}
			\includegraphics[width=\linewidth]{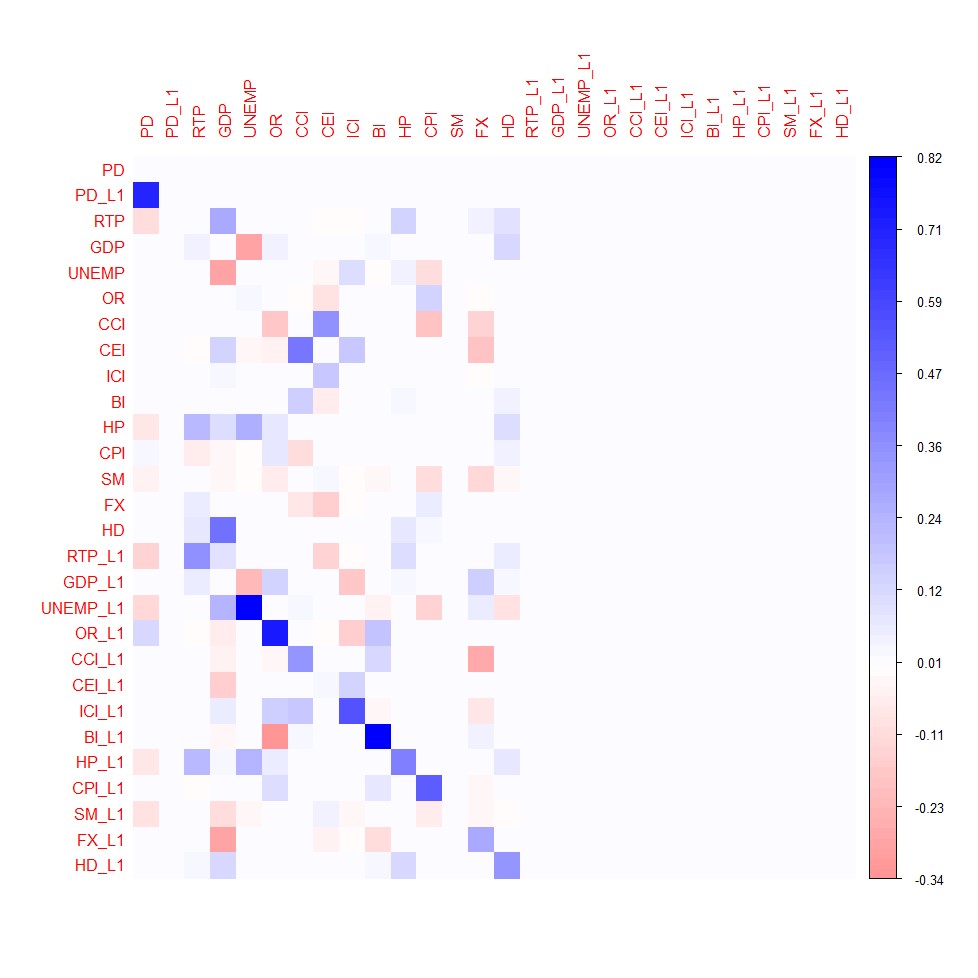}
		\end{subfigure}
    \caption{Scale of importance of macroeconomic shocks. This graph measures how sensitive the risk parameter is to the shocks of each macroeconomic variable.}
     \label{corr_partial}
	\end{figure}  

Therefore, we recommend the use of STN when selecting candidate variables for stress test models, the criterion is to prioritize the neighbors of the parameter in the network. For example, in the case study this paper, the set $\mathcal{N}e(\textrm{PD})$ make up the possible candidates. Based on our experience in multiple portfolios and risk parameters, this criterion of selection of variables gives very good results. A complementary criterion in the selection of variables is to consider the partial correlations, which is obtained previously standardized all time-series and later solving Equations (\ref{lasso}) and (\ref{elasticnet}). The partial correlations presented in the Fig. \ref{corr_partial} can be interpreted as a scale of importance of the shocks, which can be useful in the identification and control of systematic risk.


    \section{Conclusions}

In this paper, we propose to use a graphic description of the causal structure between the macroeconomic variables and the risk parameters as a useful tool for the construction of models in stress testing and thus guarantee an optimal evaluation of the resilience of risk parameters. Since this structure is underlying the data and not directly observable, we have proposed a graphical model to the reconstruction this structure. This model is quite flexible and allows to incorporate specific characteristics on this type of networks. The estimation of the model in high-dimensional contexts is carried out efficiently through regularization methods. The reconstruction of this type of networks using this methodology yields results consistent with economic and business criteria of stress testing.

	\section*{Acknowledgments}
	The authors thank Luciana Guardia for the support and motivation for this investigation to come to light. This research was supported by Santander Brasil, but the ideas and approach presented here are entirely the responsibility of the authors and not the bank.

	
	\bibliographystyle{apa}
	\bibliography{references}

\clearpage
\appendix
\section{Appendix}\label{macros}

\vspace{4cm}

\begin{center}	
\begin{tabular}{cc}				
\hline				
Variable	&	Description	\\	\hline
PD	&	Probability of Default	\\	
RTP	&	Real Total Payroll	\\	
GDP	&	Gross domestic product	\\	
UNEMP	&	Unemployment Rate	\\	
OR	&	Oficial Rate	\\	
CCI	&	Consumer Confidence Index	\\	
CEI	&	Consumer Expectation Index	\\	
ICI	&	Industry Confidence Index	\\	
BI	&	Burden Income (average of the period)	\\	
HP	&	Housing Price	\\	
CPI	&	Consumer Price Index (CPI)	\\	
SM	&	Stock Market  (en of the period)	\\	
FX	&	BRL/USD (end of period)	\\	
HD	&	Household Debt (average of the period)	\\	
\hline			
Sufix	&	Description	\\	
Variable-L1 &  Variable in Lag 1	\\	\hline
\end{tabular}	
\end{center}


\clearpage

\end{document}